\documentclass[aps,pre,twocolumn,showpacs,superscriptaddress,floatfix,10pt,amsmath,amssymb,bibnotes,footinbib,floatfix]{revtex4-2}
\usepackage{graphicx}
\usepackage{bm,bbm,bbold}
\usepackage[urlcolor=blue]{hyperref}
\hypersetup{colorlinks=true,allcolors=blue}
\usepackage{amssymb,amsfonts} 
\usepackage{color, xcolor}
\usepackage{dsfont}
\usepackage{physics}
\usepackage{amsthm}

\newcommand{\rmd}{\mathrm{d}}
\newcommand{\rme}{\mathrm{e}}

\newcommand{\f}{\bm}

\newcommand{\tf}{{t_\mathrm{f}}}
\newcommand{\lf}{{\lambda_\mathrm{f}}}

\begin{document}

\preprint{APS/123-QED}

\title{
Optimal-work feedback on particles with activity --- gliding on active fluctuations using positional information
}

\author{Lars Torbj\o rn Stutzer}\email{lars.stutzer@ds.mpg.de}
\affiliation{Max Planck Institute for Dynamics and Self-Organization, 37077 G\"ottingen, Germany}
\author{Sarah A.~M. Loos}\email{sarah.loos@ds.mpg.de}
\affiliation{Max Planck Institute for Dynamics and Self-Organization, 37077 G\"ottingen, Germany}

\date{\today}

\begin{abstract}
We study the minimum-work feedback control of particles subject to active fluctuations. Considering an active Ornstein–Uhlenbeck particle confined by a moving harmonic trap, we derive exact optimal protocols following an initial position measurement. Our results show that nonequilibrium correlations between position and active fluctuations allow work extraction from the activity based on positional information only, i.e., without directly measuring the active degree of freedom, which was the focus of earlier literature. We find that depending on the persistence time, activity can either facilitate or impede transport relative to passive systems. Surprisingly, unlike feedback schemes based on direct measurements of the active fluctuations, positional feedback remains energetically advantageous even in the limit of infinitely persistent activity. Our results provide design principles for information engines and optimal control strategies operating in active environments.
\end{abstract}

\maketitle

\section{Introduction}

The control of fluctuating microscopic systems at minimal energetic cost is a central problem in stochastic thermodynamics, with applications ranging from engineering to biology \cite{Toyabe_2012RecoveryPotential, Toyabe_2011ATPmotor, Chen2022AdvancesNanoRobots,blaber2023optimal,alvarado2026optimal}. Besides its practical relevance, this problem has also revealed fundamental connections between stochastic energetics and information theory. 
In the noise-dominated regime, information obtained from measurements can be used to dramatically reduce the work required to drive a system between prescribed states, or even enable net work extraction~\cite{Abreu2012Feedback,Abreu_2011ExtractingWork,Cao2009ThermodynamicsFeedback,Sagawa2012FeedbackControl,Sagawa2010GeneralizedJarzynski}; paving the way toward microscopic information engines.
 Such information-based work extraction is well understood for passive systems, where measurements reveal fluctuations generated by a thermal environment \cite{duBuisson2024PerformanceLimits}. A natural question is how these ideas extend to active systems, where nonequilibrium fluctuations provide an additional energetic resource and noise source.

Recent works have explored feedback control that harvest active fluctuations for work extraction~\cite{GarciaMillan_2025,Schuttler_2025,szamel2026machine,cocconi2023optimal,cocconi2024efficiency,malgaretti2022szilard,Saha2023InformationEngineNonEqBath}.
In particular, using analytically tractable models of single harmonically trapped active particles, the fundamental performance limits to work extraction by adapting the trap stiffness or trap position have been established using optimal control theory~\cite{GarciaMillan_2025,Schuttler_2025,szamel2026machine,cocconi2023optimal,cocconi2024efficiency,olsen2025harnessing,goswami2025optimal}. 
However, these earlier works share in common that the feedback is based on direct measurements of the active degree of freedom itself, such as the direction and magnitude of self-propulsion. 

In many situations, however, the active degree of freedom is not directly observable. It may represent an internal propulsion state that is difficult to access experimentally---for example the configuration of bacterial flagella or the local chemical state surrounding a catalytic Janus particle.
An equally important application concerns passive tracers immersed in active baths, such as suspensions of synthetic active particles~\cite{Boriskovsky2026EffectiveTemp, Boriskovsky2024FDTactivebath} or bacterial baths~\cite{Maggi2017FDTactivebath, Chen2007RheologyActiveBath, VillalobosConcha2025ActiveBathDroplets, kushwaha2026unconventionalgrowthkineticsfractal}. Here, the active fluctuations represent (in a coarsegrained way) the collective  effect of many hidden bath degrees of freedom, rendering direct measurements impossible. By contrast, the particle position is typically directly measurable. 

This raises a basic question: can active fluctuations be harnessed using positional information alone? Recent studies on information engines suggest that positional information may indeed be sufficient to extract work from activity~\cite{malgaretti2022szilard,Saha2023InformationEngineNonEqBath}. However, the fundamental performance limits of such measurement schemes and the corresponding optimal control strategies remain elusive. 

Here, we address this question using  a minimal, analytically tractable model of active Ornstein–Uhlenbeck particle confined by a moving harmonic trap. The model can describe both, an intrinsically active particle in a thermal enviornment, or a passive particle driven by an active environment. 
The key observation is that 
the experimentally accessible positional information is enough to efficiently harvest the hidden activity---thanks to the nontrivial correlations between position and active fluctuations that are inherent to the nonequilibrium system, even in the steady state.  
In some regimes, positional optimal feedback can even outperform direct measurements of the active fluctuations. For example, whereas direct measurements lose their energetic advantage for large persistence times, positional feedback remains beneficial even in the limit of infinitely persistent activity.

The remainder of this paper is organized as follows. We first state the general control objective, define the model, and describe how measurements infer information about the different degrees of freedom. Subsequently, we derive the optimal solution for initial position measurement and compare the performance of different control schemes, with and without feedback. Finally, we analytically investigate the asymptotic behaviors, which are then mapped onto passive systems to gain a physical intuition. We conclude at the end.

\section{Model and control problem}
\subsection{Control objective}\label{sec:objective}
We consider the optimal control problem of translating a trapping potential $V$ which has a minimum at $\lambda(s)$ and contains a particle from a prescribed initial position $\lambda_0\equiv\lambda(0)=0$ to a prescribed final value $\lambda_\mathrm{f}\equiv\lambda(t_\mathrm{f})$ in a given time $t_\mathrm{f}$ using minimum average work input
\begin{align}
    W_\alpha[\lambda] = \left\langle \int_{s=0}^{s=t_\mathrm{f}}
    \frac{\partial V}{\partial \lambda} \circ\rmd\lambda(s)\right\rangle_\alpha\,.\label{MotivationWork}
\end{align}
Here, $\circ \, \rmd \lambda$ denotes a Stratonovich integral and $\langle\cdot\rangle_\alpha$ is the expectation over all stochastic degrees of freedom, conditioned on a specific initial system state, indicated by $\alpha$, at the onset of the control. Choosing the conditioning appropriately allows us to investigate the effect of initial measurements.
Before any control action or measurement are done, we assume that the process starts in a (nonequilibrium) steady state. 
Choosing an initial condition from the unperturbed steady state therefore corresponds to  \textit{feedforward} (\textit{open-loop}) control. On the contrary, specifying an initial condition that corresponds to the probability distribution that describes the state after a measurement is done (specified below) allows us to model \textit{feedback} (\textit{closed-loop}) control. 
For passive particles this problem was thoroughly investigated in \cite{Abreu_2011ExtractingWork}. For active particles, this problem was recently investigated in \cite{Schuttler_2025,GarciaMillan_2025} with a focus on direct measurements of the active fluctuations.

A central objective of this paper is to
understand the influence of $\alpha$ on the optimal solutions of Eq.~\eqref{MotivationWork} 
and investigate the effect of activity entering through the fluctuations of the particle. We will thereby focus on measurements of the positional degree of freedom.

\subsection{Active Ornstein Uhlenbeck particles}\label{sec:AOUP}
Let $x(s)\in\mathbb{R}$ denote the position of an AOUP at time $0\leq s\leq \tf$ obeying the one-dimensional overdamped Langevin equation \cite{Martin2021MechanicsAOUP, Bonilla_2019AOUP} 
\begin{align}
    \rmd x(s) &= -\kappa[x(s)-\lambda(s)]\rmd s + v(s)\rmd s + \sigma \rmd B_s\,,\label{Langevin}\\
    \tau\,\rmd v(s) &= -v(s)\rmd s + \sigma' \rmd B_s'.\label{AOU}
\end{align}
The drift term in Eq.~\eqref{Langevin} is due to a harmonic potential $V(x) = \frac{\kappa}{2}[x-\lambda]^2$, realizable in experiments, e.g., via an optical trap with strength $\kappa$ and centered at position $\lambda$. The random displacement $\sigma\rmd B_s$ models standard thermal fluctuations due to coupling of the particle to a heat bath with $\langle \rmd B_s\rangle=0$ and $\langle \rmd B_s\rmd B_z\rangle = \delta(s-z)\rmd s\rmd z$. The term $v(s)$ in Eq.~\eqref{Langevin} breaks detailed balance and introduces \textit{activity}, which could either come from an intrinsic self-propulsion mechanism of the particle (then $x$ models the motion of an active particle in a passive bath), or from active fluctuations in the surrounding bath (then $x$ models a passive particle in an active bath). Importantly, at the level of abstract mathematical description, we do not need to distinguish between the two scenarios so that our results apply to both.
Equation~\eqref{AOU} describes the evolution of the resulting active fluctuations $v(s)$ on $x$, also called ``active velocity,'' which have a persistence time $\tau$ [that is the timescale on which 
$v(s)$
decorrelates], 

where $\rmd B_s'$ is another Wiener process with the same properties as $B_s$ that is uncorrelated with $B_s$. 

Due to the linearity of \eqref{Langevin} and \eqref{AOU}, the joint stationary distribution of $\f Y=(x(0),v(0))\equiv(x_0, v_0)\sim\mathcal{N}(\f\mu, \f\Sigma)$ inherits the Gaussianity of the noises, and is characterized by mean $\f\mu=\f0$ and covariance matrix $\f \Sigma$ with entries \cite{Schuttler_2025, GarciaMillan_2025, szamel2026machine}
\begin{subequations}\label{covariances}
\begin{align}
    \mathrm{var}(x_0) &=\frac{1}{\kappa}\left(D+\frac{\omega^2\tau}{\kappa\tau+1}\right)\,,\label{varx}\\
    \mathrm{cov}(x_0, v_0)&=\frac{\omega^2\tau}{\kappa\tau+1}\,,\\ 
    \mathrm{var}(v_0)&=\omega^2\,,
\end{align}
\end{subequations}
where $D=\sigma^2/2$ is the diffusion coefficient of $x$ and $\omega^2\equiv{\sigma'}^2/2\tau$ scales the variance of the active fluctuations.

To ensure ergodicity for all $\tau$, we set
$\omega^2=\mathrm{const.}$, where both asymptotic limits $\tau\to0$ and $\tau\to\infty$ have well-defined stationary distributions. The limit $\tau\to0$ of Eq.~\eqref{Langevin} recovers passive Ornstein-Uhlenbeck dynamics, because Eq.~\eqref{AOU} reduces to $v(s)\rmd s=0$, so that the setting corresponds to the control of an overdamped passive particle~\cite{Schmiedl2007,Abreu_2011ExtractingWork}. The limit $\tau\to\infty$ again reduces Eq.~\eqref{Langevin} to a passive Ornstein-Uhlenbeck process but with an additional constant drift whose value is randomly set by the initial condition $v_0$. 
Other possible scalings of $\sigma'$  are discussed in Appendix~\ref{AppendixScaling}.

\subsection{Positional feedback of active particles}\label{sec:InferredInfo}

The open-loop control problem described above has already been solved in Refs.~\cite{Schuttler_2025, GarciaMillan_2025}. The essential insight is that the optimal protocols are \textit{identical} to the case of the corresponding passive case, i.e., a simple Brownian (passive) particle. The only effect of activity is an increase in the work fluctuations. In contrast, including a measurement operation on the stochastic system before deciding on the protocol has been shown to lead to nontrivial optimal protocols with reversed jumps and, under suitable conditions, allow for net work extraction ($W_\alpha<0$). Specifically, Refs.~\cite{Schuttler_2025, GarciaMillan_2025} discuss the effect of an initial measurement of $v$, which requires knowing both the orientation (sign) and strength (magnitude) of the active fluctuations. However, this type of information is often difficult to access in experiments, as it typically requires knowledge of underlying chemical or structural states of hidden variables. Moreover, for a passive particle in an active bath where $v(s)$ is a coarsegrained variable that describes the internal state of the bath degrees of freedom, $v_0$ might become fully inaccessible; a case which is thus not discussed in Refs.~\cite{Schuttler_2025, GarciaMillan_2025}.

Here, we will instead study closed-loop control after measured (or fixed) initial \textit{position}, $\alpha=x_0$, which lends itself more directly to experimental realizations~\cite{Liu2002VelocityMeasurement, Brown1992AlgorithmDesign}, and allows us to explicitly include the case of a passive particle in an active bath.

\subsection{Underlying physical mechanisms of work extraction and expectations}

While transporting a particle through a medium generally requires work input, utilizing information about the fluctuating system state can turn the average work negative. There are two known mechanisms of work extraction that are expected to play a role for the problem at hand. 

First, the persistent motion in a given direction means that the particles behave, on short timescales, akin to ballistic particles with inertia (and kinetic energy). This allows work to be harvested from the activity itself by moving the potential along the persistent motion, which leads the particle to persistently climb up the potential. Such extraction mechanism requires a finite amount of extraction time and can last until the persistent motion has lost it's direction due to noise. Intuitively, one might expect that $x$-information is not, or at least less, useful to harvest work on this route from activity itself, as it does not probe the active degree of freedom directly. We will investigate this question by comparing measurements of the position versus the active fluctuation.
The key insight is that, because the particle position and active fluctuations are nontrivially correlated in such nonequilibrium system, work can in fact be efficiently harvested from the underlying activity by using positional information alone.

Secondly, one should recall that, even for a passive system, positional feedback can lead to work extraction from (thermal) fluctuations. The reason is that thermal fluctuations transiently increase the potential energy in the trap, which can be harvested by moving the trap to release the excitations~\cite{Abreu_2011ExtractingWork}. Such extraction mechanism happens by an instantaneous repositioning of the trap.
Since activity increases positional fluctuations (by increasing the ``effective temperature"~\cite{Loi_2008EffectiveTemperatureActiveMatter, Hecht2024DefineActiveTemperature,wiese2024modeling}), one might expect that this mechanism of work extraction is enhanced by activity.
The question that now emerges is to what extent activity really promotes (or hinders) the conversion of excitations of potential energy into work; and how by the combination of both mechanisms, the work extraction after positional and after active-fluctuation measurements compare.

\section{Optimal closed-loop solutions}

\subsection{Expectation values and Correlations}
Before computing the optimal solutions, it is useful to derive the expectation and average values of the observable and non-observable variables after a positional measurement. In the remaining parts of the manuscript, we denote $\langle\cdot\rangle$ as the expectation over all noise histories and $\mathbb{E}_\beta[\cdot|\alpha]$ the expectation over all values of $\beta$ conditioned on $\alpha$. Specifically, since we consider fixed initial positions, the inferred value of $v$ from a positional measurement (i.e., conditioned on {$x_0$}) is
\begin{align}
    v_{x_0}\equiv \mathbb{E}_{v_0}[v_0|x_0] = \frac{\mathrm{cov}(x_0, v_0)}{\mathrm{var}(x_0)}x_0\,,\label{vgivenx}
\end{align}
where we use the fact that the joint stationary distribution is Gaussian.  
Using Eqs.~\eqref{covariances},  Eq.~\eqref{vgivenx} can be evaluated, leading to
\begin{align}
    v_{x_0} = \frac{\omega^2\kappa\tau}{D(\kappa\tau + 1) + \omega^2\tau}x_0\,.\label{vConditionedX}
\end{align}
This quantity reflects the nontrivial correlations between positional and active fluctuations that are inherent to the nonequilibrium steady state. 
We will in the following show how these fluctuations can be exploited to glide on active fluctuations, i.e., move without work input, or even extract work from activity while moving the trap.

By solving Eq.~\eqref{AOU} and inserting the inferred initial value for $v_0$, we can further derive the average time evolution of the active velocity conditioned on the initial position
\begin{align}
    u_{x_0}(s)&\equiv \langle v(s)\rangle_{x_0} = \mathbb{E}_{v_0, x_0}\left[\langle v(s)\rangle|x_0\right]= v_{x_0}\mathrm{e}^{-\frac{s}{\tau}}\,,\label{u_x0}
\end{align}
while the conditioned average position $q_{x_0}(s)\equiv \langle x(s)\rangle_{x_0}=\mathbb{E}_{v_0, x_0}\left[\langle x(s)\rangle|x_0\right]$ from Eq.~\eqref{Langevin} 
follows the evolution equation
\begin{align}
    \dot q_{x_0}(s) = -\kappa[q_{x_0}(s) - \lambda(s)] + u_{x_0}(s)\,.\label{q_x0}
\end{align}

Moreover, due to the Gaussian nature of the system, we can calculate the conditional variances of the system conditioned on $x$ or $v$ measurements, for which we find
\begin{align}
    \!\Sigma_{x_0x_0|v_0} &= \Sigma_{x_0x_0} - \frac{\Sigma_{x_0v_0}^2}{\Sigma_{v_0v_0}} = \frac{D(\kappa\tau+1)^2+\omega^2\tau}{\kappa(\kappa\tau+1)^2}\,,\\
    \!\Sigma_{v_0v_0|x_0} &= \Sigma_{v_0v_0} - \frac{\Sigma_{x_0v_0}^2}{\Sigma_{x_0x_0}}=\frac{D\omega^2(\kappa\tau+1)^2 + \omega^4\tau}{D(\kappa\tau+1)^2+\omega^2\tau(\kappa\tau+1)}\,,
\end{align}
where the double indices denote (co-)variances, e.g., $\Sigma_{x_0x_0}\equiv\mathrm{var}(x_0)$  and $\Sigma_{x_0 x_0|v_0}$ is the variance of the conditional probability $P(x_0|v_0)$. The conditional variances will be useful to quantify the typical fluctuation of $v_{x_0}$ (or $x_{v_0}$) upon a measurement of $x_0$ (or $v_0$, respectively), allowing us to systematically compare the influence of measurements. 

\begin{figure}
    \centering
    \includegraphics[width=\linewidth]{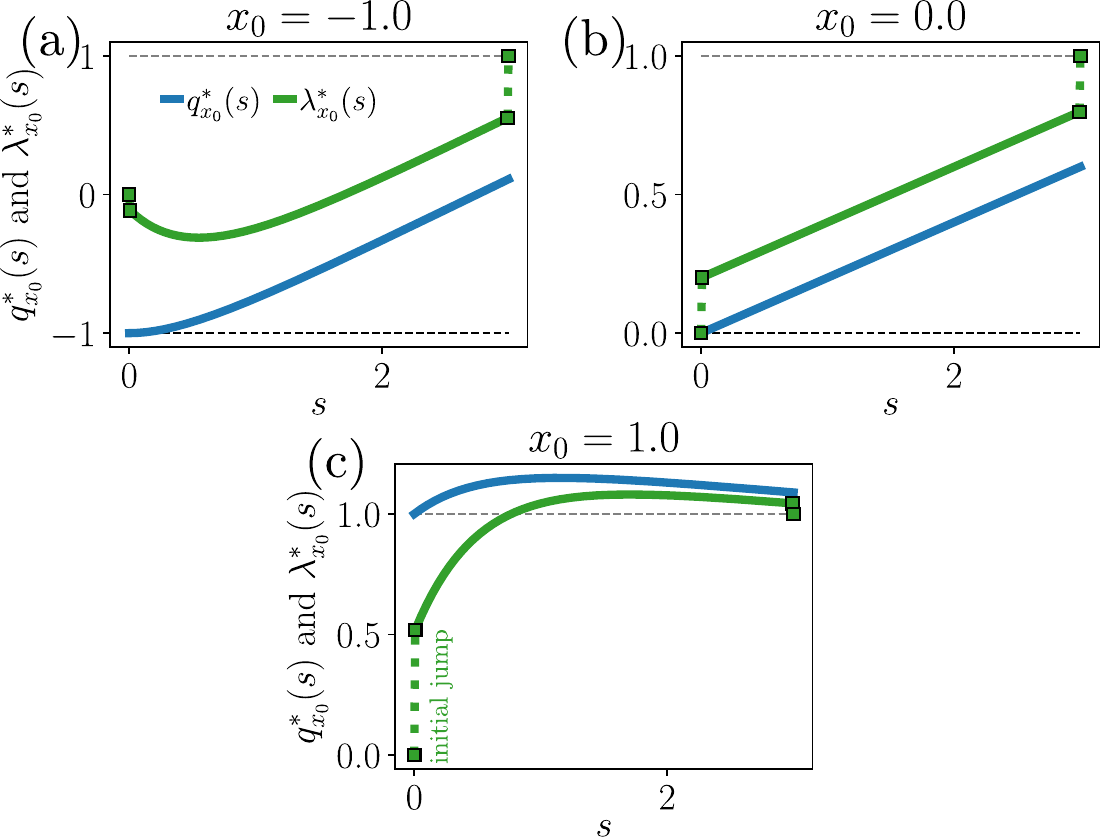}
    \caption{ 
    Examples of optimal protocols (green) and particle trajectories (blue) [Eqs.~\eqref{optimalSolution}] for different initial values $x_0$ (black dashed lines). The green dashed lines at the beginning and end of the protocol show the instantaneous jumps. Here, the parameters are set to $\lf=1$ (gray dashed lines), $\tf=3$, $\tau=0.5$, $\kappa=1$, $D=1$, and $\omega=5$.}
    \label{fig:fig1}
\end{figure}

\begin{figure*}
    \centering
    \includegraphics[width=.8\linewidth]{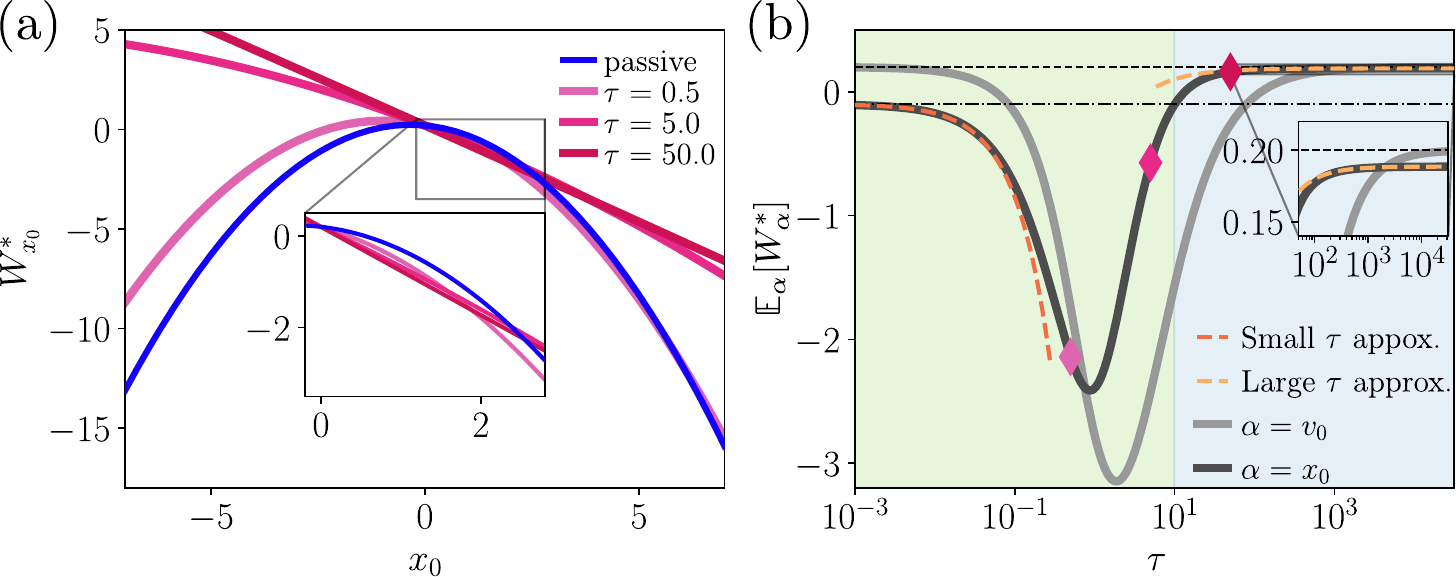}
    \caption{ Influence of $\tau$ on the optimal work. 
    (a) Optimal work $W_{x_0}^*$ after positional measurement as a function of $x_0$ for different values of $\tau$. The inset shows a zoom-in on the region where an increase in $\tau$ lowers the required work needed to displace the trap. (b) Work after positional measurements (black) and after direct measurement of active fluctuations (gray) as functions of $\tau$. Dashed plots show the asymptotic expansions for small $\tau$ [dashed orange,  Eq.~\eqref{smallTau}] and large $\tau$ [dashed yellow, Eq.~\eqref{largeTauCorrection}]. 
    The dash-dotted horizontal line shows the measurement-averaged work after positional measurements in the passive case, $\mathbb{E}_{x_0}^\mathrm{pas}[W_{x_0, \mathrm{pas}}^*]$, while the dashed horizontal line shows the open-loop optimal work, $W_\mathrm{ff}^*=\kappa\lambda_\mathrm{f}^2/(\kappa\tf + 2)$. Accordingly, in the green shaded region activity lowers the work, $\mathcal W^\mathrm{ex}_\mathrm{act-pas,x_0}<0$, while in the blue shaded region activity increases the required work input, $\mathcal W^\mathrm{ex}_\mathrm{act-pas,x_0}>0$, but measurements are still beneficial, $\mathcal W^\mathrm{ex}_{\mathrm{fb-ff,x_0}}<0$. The inset shows how the work with direct active-fluctuation feedback converges to the feedforward value $W_\mathrm{ff}^*$ (dashed line) while for positional feedback the work stays below it. 
    The remaining parameters are: $\tf=3$, $\lf=1$, $\kappa=1$, $D=1$,
    and $\omega=5$.}
    \label{fig:fig2}
\end{figure*}

\subsection{Optimal solutions}

We now have all ingredients to explicitly derive the protocols $\lambda(t)$ that optimize the work required for the transport problem introduced in Sec.~\ref{sec:objective} with  boundary conditions $\lambda(0)=0$ and $\lambda(\tf)= \lf$, after an initial position measurement, $\alpha=x_0$. The corresponding mean work functional [defined in Eq.~\eqref{MotivationWork}] reads
\begin{align}
    W_{x_0}[\lambda] = \kappa\int_0^{t_\mathrm{f}}\rmd s \dot\lambda(s)\left[\lambda(s) - q_{x_0}(s)\right]\,.
\end{align}
By performing an integration by parts, inserting Eq.~\eqref{q_x0}, and solving the variational problem, the optimal trap protocol, average colloid trajectory, and work are found to be (for a detailed derivation, see Appendix~\ref{AppendixDerivation}) \cite{Schuttler_2025, GarciaMillan_2025}
\begin{align}
    \lambda^*_{x_0}(s) &= x_0+\frac{\kappa s+1}{\kappa t_\mathrm{f}+2}d_{x_0} + \frac{v_{x_0}}{2}\left[\tau-\left(\tau+\frac{1}{\kappa}\right)\mathrm{e}^{-s/\tau}\right]\,,\nonumber\\
    q_{x_0}(s)&=x_0 + \frac{\kappa s}{\kappa t_\mathrm{f} + 2}d_{x_0} + \frac{\tau}{2}v_{x_0}\left(1-\mathrm{e}^{-s/\tau}\right)\,,\nonumber\\
    W_{x_0}^*&=\frac{\kappa}{\kappa t_\mathrm{f}+2}d_{x_0}^2 - \frac{\kappa}{2}x_0^2 - \frac{\tau}{8}v_{x_0}^2\left(1-\mathrm{e}^{-2t_\mathrm{f}/\tau}\right).\label{optimalSolution}
\end{align}
Here and in the following, we use the superscript * to mark optimal solutions. 
Furthermore, $d_{x_0}$ denotes an effective distance given by 
\begin{align}
    d_{x_0}\equiv \lambda_\mathrm{f} - x_0 - \frac{\tau}{2}v_{x_0}\left(1-\mathrm{e}^{-t_\mathrm{f}/\tau}\right)\,.
\end{align}
The optimal trajectories in Eqs.~\eqref{optimalSolution} are illustrated in Figs.~\ref{fig:fig1}(a)-(c), for three exemplary values of $x_0\in\{-1, 0, 1\}$.
The optimal protocols exhibit generic discontinuities at $s=0$ and $s=\tf$, as expected from the passive case. While for feedforward control schemes, these jumps are known to be symmetric and always pointing towards to final position $\lf$ [see Fig.~\ref{fig:fig1}(b)] \cite{Schmiedl2007,Loos_2024}, any nonzero measured $x_0$ leads to asymmetric and possibly reversed jumps, see Fig.~\ref{fig:fig1}(a) and~(c) for examples, consistent with earlier observations for feedback after $v$-measurements~\cite{GarciaMillan_2025}. Furthermore, when $x_0\neq 0$, the optimal protocol and mean particle trajectory are generally nonlinear between the jumps, due to the exponential decorrelation of active fluctuations.
For the specific measurement outcome of $x_0=0$, the solutions reduce to the corresponding solutions of feedforward control, and loose any trace of the underlying activity [see Fig.~\ref{fig:fig1}(b)].

To gain further intuition, consider Fig.~\ref{fig:fig2}(a) showing $W_{x_0}^*$ as functions of $x_0$, for different $\tau$. 

The asymmetry of the curves indicates that it is energetically less costly, if the orientation of the active fluctuations is in the target direction (here, $\lf=1$), as expected. The active fluctuations then effectively transport the particle in the target direction,
so that less work is needed than the passive counterpart shown in blue ($W_{x_0}^*<W_{x_0,\mathrm{pas}}^*$).
On the other hand, initial positions away from the target, $x_0<0$ (corresponding to net active fluctuations opposing the target direction of the trap $v_{x_0}\propto x_0<0$), requires additional energetic input to steer the particle towards $\lf$ against the active motion, generally resulting in higher required work ($W_{x_0}^*>W_{x_0,\mathrm{pas}}^*$). The same happens 
in the regime of too large active velocities, or too strong persistence, which lead to overshooting the target within $\tf$, so that \textit{counter-steering} becomes necessary.
Thus, increasing $\tau$ decreases the range of beneficial $x_0$ values.

The optimal work in Eq.~\eqref{optimalSolution}, while noise-averaged, still 
fluctuates with the initial condition (measurement outcome) $x_0\sim \mathcal{N}(0, \Sigma_{x_0x_0})$. To access the mean benefit of position measurements, we therefore need to take the weighted average over all realizations of $x_0$. 
Using Eq.~\eqref{varx} and Eq.~\eqref{vConditionedX}, we obtain the optimal work averaged over initial positions
\begin{widetext}
    \begin{align}
     \mathbb{E}_{x_0}\left[W_{x_0}^*\right]=&\frac{\kappa}{\kappa t_\mathrm{f}+2}\lambda_\mathrm{f}^2 + \frac{\kappa}{\kappa t_\mathrm{f}+2}\mathbb{E}_{x_0}\left[\left[x_0 + \frac{\tau}{2}v_{x_0}\left(1-\mathrm{e}^{-t_\mathrm{f}/\tau}\right)\right]^2\right] - \frac{\kappa}{2}\mathbb{E}_{x_0}\left[x_0^2\right] - \frac{\tau}{8}\mathbb{E}_{x_0}\left[v_{x_0}^2\right]\left(1-\mathrm{e}^{-2t_\mathrm{f}/\tau}\right)\nonumber\\
     =&\frac{\kappa}{\kappa t_\mathrm{f}+2}\lambda_\mathrm{f}^2 +\left\{ \frac{1}{\kappa t_\mathrm{f}+2}\left[1 + \frac{\omega^2\kappa\tau^2}{2D(\kappa\tau + 1) + 2\omega^2\tau}\left(1-\mathrm{e}^{-t_\mathrm{f}/\tau}\right)\right]^2 -\frac{1}{2} \right.
     \nonumber\\&
     -\left.\frac{\omega^4\kappa\tau^3}{8\left[D(\kappa\tau + 1) + \omega^2\tau\right]^2}\left(1-\mathrm{e}^{-2t_\mathrm{f}/\tau}\right)\right\}\left(D + \frac{\omega^2\tau}{\kappa \tau + 1}\right)\,.\label{OptimalWork}
    \end{align} 
\end{widetext}
The result of Eq.~\eqref{OptimalWork} is presented in Fig.~\ref{fig:fig2}(b) for an exemplary parameter choice. For comparison, the dashed line shows the open-loop optimal work, $W^*_\mathrm{ff} \equiv \kappa\lambda_{\mathrm{f}}^2/(\kappa\tf+2)$ \cite{Schmiedl2007}, which is found to be strictly above the average extracted work for optimal positional feedback in the active case. Note that $W^*_\mathrm{ff}$ is identical for active and passive systems~\cite{Schuttler_2025, GarciaMillan_2025, Schmiedl2007, GomezMarin_2008}. Furthermore, the dash-dotted line shows the optimal positional feedback work in a passive system, $\mathbb{E}_{x_0}^\mathrm{pas}\left[W_{x_0, \mathrm{pas}}\right]$, obtained by taking the limit $\tau\to 0$ of Eq.~\eqref{optimalSolution}, which yields
\begin{align}
    W_{x_0, \mathrm{pas}}^* &= \frac{\kappa}{\kappa t_\mathrm{f}+2}(\lambda_\mathrm{f}-x_0)^2 - \frac{\kappa}{2}x_0^2\,,\label{PassiveFBwork}\\
    \mathbb{E}_{x_0}^\mathrm{pas}\left[W_{x_0, \mathrm{pas}}^*\right]&=\frac{\kappa}{\kappa t_\mathrm{f}+2}\lambda_\mathrm{f}^2 - \frac{\kappa^2 t_\mathrm{f}}{2(\kappa t_\mathrm{f}+2)}\mathbb{E}_{x_0}^\mathrm{pas}[x^2_0]\nonumber\\
    &= \frac{\kappa}{\kappa t_\mathrm{f}+2}\lambda_\mathrm{f}^2 - \frac{\kappa t_\mathrm{f}D}{2(\kappa t_\mathrm{f}+2)}\,,\label{PassiveFBworkAverage}
\end{align}
where we have used that $\mathbb{E}_{x_0}^\mathrm{pas}[x^2_0]=D/\kappa$. The shaded regions in Fig.~\ref{fig:fig2}(b) correspond to $\mathbb E_{x_0}\left[W_{x_0}^*\right]\leq\mathbb{E}_{x_0}^\mathrm{pas}\left[W_{x_0, \mathrm{pas}}^*\right]$ (green) and $\mathbb E_{x_0}\left[W_{x_0}^*\right]\geq\mathbb{E}_{x_0}^\mathrm{pas}\left[W_{x_0, \mathrm{pas}}^*\right]$ (blue), depicting when active fluctuations result in more or less costly optimal transport compared to the passive closed-loop counterpart. 

We find that for small $\tau$ the average work is decreased compared to the passive positional feedback work [green shaded region in Fig.~\ref{fig:fig2}(b)]. This means that, even when averaged over all possible measurement outcomes, the trap can effectively harvest energy from the activity by gliding along the active fluctuations. This mechanism benefits from increasing $\tau$ which extends the effective harvesting time, before the persisting active fluctuations decorrelate (i.e., become zero on average).
However, beyond a certain point, increasing $\tau$ results, on average, in the particle overshooting the target, which causes the work to eventually increase with $\tau$, leading to a minimum in the work at the onset of this effect. Ultimately, this overshooting and counter-steering lets the work become even larger than for the passive case (blue shaded region).

Finally, we aim to compare the energetic benefit of the two different measurements ($x_0$ versus $v$). Comparing the mean work after positional and active-fluctuation measurements [black versus grey line in Fig.~\ref{fig:fig2}(b)], we see that the lines intersect multiple times and have different limiting behaviors. Beyond plotting the curves for a specific parameter choice, as done in Fig.~\ref{fig:fig2}(b), it is however hard to read off the general trends based on the analytical expressions, which are very cumbersome [see Eq.~\eqref{OptimalWork}]. However, we can perform an asymptotic analysis, where the expressions simplify. 

To this end, we define two measures of \emph{excess work}: one relative to the corresponding feedforward control [dashed line in Fig.~\ref{fig:fig2}(b)] and another relative to positional-feedback control of a passive particle [dash-dotted line in Fig.~\ref{fig:fig2}(b)], given by
\begin{align}
    \mathcal W^\mathrm{ex}_{\mathrm{fb-ff,x_0}} &\equiv \mathbb{E}_{x_0}\left[W_{x_0}^*\right] - W^*_\mathrm{ff}\,,\label{excessFFworkx0}\\
    \mathcal W^\mathrm{ex}_\mathrm{act-pas,x_0}&\equiv \mathbb{E}_{x_0}\left[W_{x_0}^*\right] - \mathbb{E}_{x_0}^\mathrm{pas}\left[W_{x_0,\mathrm{pas}}^*\right]\,,\label{excessWork}
\end{align}
respectively.
 Hence, the two questions---When are $x_0$-measurements and activity beneficial?---reduce to under what conditions $\mathcal W^\mathrm{ex}_{\mathrm{fb-ff,x_0}}\leq 0$ and $\mathcal W^\mathrm{ex}_\mathrm{act-pas,x_0}\leq 0$ hold, respectively.
To compare with control after measuring the active velocity [grey line in Fig.~\ref{fig:fig2}(b)], we analogously define the excess work due to $v$-measurement
\begin{align}
     \mathcal W^\mathrm{ex}_{\mathrm{fb-ff,v_0}} \equiv\mathbb{E}_{v_0}\left[W_{v_0}^*\right] - W_\mathrm{ff}^*\,.
\end{align}
Note that due to the lack of an appropriate passive counterpart, we cannot define an %
excess work due to activity for closed-loop control with $v$-measurements.

Now we consider the asymptotic regimes of small and large persistence times, where the expressions become more tractable. 

We start with the limit $\tau\to 0$ of Eq.~\eqref{excessWork},
\begin{align}
    \!\!\lim_{\tau\to 0}\mathcal W^\mathrm{ex}_\mathrm{act-pas,x_0} = \lim_{\tau\to 0}\mathbb{E}_{x_0}\left[W_{x_0}^*\right] - \mathbb{E}_{x_0}^\mathrm{pas}\left[W_{x_0,\mathrm{pas}}^*\right],
\end{align}
where we find
\begin{align}
    \lim_{\tau\to 0}\mathbb{E}_{x_0}\left[W_{x_0}^*\right] = \frac{\kappa}{\kappa t_\mathrm{f} + 2}\lambda_\mathrm{f}^2 + \left\{\frac{\kappa}{\kappa t_\mathrm{f} + 2} - \frac{\kappa}{2}\right\}\frac{D}{\kappa}\,,
\end{align}
which, together with Eq.~\eqref{PassiveFBworkAverage}, exactly yields $\lim_{\tau \to 0}\mathcal W^\mathrm{ex}_\mathrm{act-pas,x_0} = 0$. This verifies, as a consistency check, that the excess work due to activity vanishes in the passive limit $\tau\to 0$. Around this limit, we can further expanded $\mathcal W^\mathrm{ex}_\mathrm{act-pas,x_0}$ in small $\tau$  (small compared to the remaining timescales of the system, $\tf$ and $1/\kappa$), yielding
\begin{align}
    \mathcal W^\mathrm{ex}_\mathrm{act-pas,x_0}&=-\frac{\kappa t_\mathrm{f}\omega^2}{2(\kappa t_\mathrm{f}+2)}\tau +\mathcal{O}(\tau^2)\,.\label{smallTau}
\end{align}
This asymptotic expression indicates that, in the regime of small persistence times, we generally benefit from activity ($\mathcal W^\mathrm{ex}_\mathrm{act-pas,x_0}\leq0$) after a positional measurement. This result also immediately implies that the excess work compared to feedforward control is always negative, $\mathcal W^\mathrm{ex}_{\mathrm{fb-ff,x_0}}\leq0$, for small $\tau$.

The situation changes qualitatively, as soon as the persistence time becomes large compared to other timescales of the dynamics. To analytically investigate the case $\tau\gg\tf$, we now consider the opposite extreme limit of infinitely persistent motion, $\tau\to\infty$, where we find
\begin{align}       \!\!\lim_{\tau\to\infty}\mathbb{E}_{x_0}\left[W_{x_0}^*\right] = \frac{\kappa\lambda_\mathrm{f}^2}{\kappa t_\mathrm{f}+2} - \frac{D^2\kappa^2 t_\mathrm{f}}{2(\kappa t_\mathrm{f}+2)(D\kappa + \omega^2)}\,,\label{WorkTauInfty}
\end{align}
and therefore
\begin{align}
    \lim_{\tau\to\infty}\mathcal W^\mathrm{ex}_\mathrm{act-pas,x_0} &= - \frac{D^2\kappa^2 t_\mathrm{f}}{2(\kappa t_\mathrm{f}+2)(D\kappa + \omega^2)} + \frac{D\kappa t_\mathrm{f}}{2(\kappa t_\mathrm{f} +2)}\nonumber\\
    &=\frac{D\kappa\omega^2 t_\mathrm{f}}{2(\kappa t_\mathrm{f}+2)(D\kappa + \omega^2)}\geq 0\,.\label{ExcessWorkTauInfty}
\end{align}

By including the leading order correction in $\tau$ in $\mathcal W^\mathrm{ex}_\mathrm{act-pas,x_0}$, which is of order $\mathcal{O}(\tau^{-1})$, we see that
\begin{align}
    \mathcal W^\mathrm{ex}_\mathrm{act-pas,x_0} &= \frac{D\kappa\omega^2\tf}{2(\kappa\tf+2)(D\kappa+\omega^2)}
    \times
    \nonumber \\
   & \left[1 -\frac{1}{\tau}\frac{D+D\kappa\tf+\tf\omega^2}{D\kappa+\omega^2} \right] + \mathcal{O}(\tau^{-2}),\label{largeTauCorrection}
\end{align}
showing a convergence to Eq.~\eqref{ExcessWorkTauInfty} from below. 
The asymptotic in Eq.~\eqref{largeTauCorrection} shows that sufficiently large persistence combined with a shorter protocol duration leads to positive excess feedback work $\mathcal{W}_\mathrm{act-pas,x_0}^\mathrm{ex}$, i.e., activity increases the work input needed to displace the trap. This is because (except for very specific initial positions) the control eventually has to counteract the persistent motion to stay on target, requiring additional energy input [see Fig.~\ref{fig:fig2}(a)].  However, $\lim_{\tau\to\infty} \mathcal W^\mathrm{ex}_{\mathrm{fb-ff,x_0}}\leq0$, meaning that, even in the limit of infinitely persistent active fluctuations, positional feedback lowers the required work input; contrasting $v$-feedback, which ceases to bring energetic benefit in this limit. Namely, as reported in Refs.~\cite{GarciaMillan_2025, Schuttler_2025},
\begin{align}
    \lim_{\tau\to\infty}\mathbb{E}_{v_0}[W_{v_0}^*] &= \mathbb{E}_{v_0}\!\left[\frac{\kappa}{\kappa\tf+2}\lambda_\mathrm{f}^2 - \lf v_0\right] 
    \nonumber \\
    &= \frac{\kappa}{\kappa\tf+2}\lambda_\mathrm{f}^2\geq 0\,.\label{v0WorkTauInfty}
\end{align}

The difference in the asymptotic $\tau\to\infty$ values can be rationalized by a mapping the active control problem to a passive control problem with a constant, but random tilt.
We give the detailed calculation and explanation in Appendix~\ref{passiveDrift} and here only summarize the main arguments. Namely, in the ballistic limit, $v$-measurements [Eq.~\eqref{v0WorkTauInfty}] lead to a less precise estimate of the ``energetic excitation'' in the form of drift and raised potential energy due to active and thermal fluctuations, compared to $x_0$ measurements [Eq.~\eqref{WorkTauInfty}]. Asymptotically, the information of $v_0$ even yields no energetic benefit. This can be understood because measuring the random tilt of a potential in a passive system, the optimal protocol effectively reduces to the non-tilted problem \cite{Schmiedl2007}.

Additionally, Fig.~\ref{fig:fig2}(b) 
shows that the type of initial measurement that results in lowest optimal transport cost varies for increasing $\tau$, demonstrating that exact information about the initial excitations is not always energetically optimal.

\section{Discussion and Outlook}
We have investigated the minimum-work feedback control of particles driven by active Ornstein–Uhlenbeck fluctuations using only positional information. Despite not directly measuring the active fluctuations, which was the focus of most of the earlier literature, we showed that the nonequilibrium correlations between position and hidden active degrees of freedom suffices for effective work extraction. This demonstrates that positional information alone is sufficient to harness activity as an energetic resource.

Comparing active and passive systems, we provided asymptotic arguments to show that activity has a dual role. For sufficiently short persistence times, active fluctuations assist transport and reduce the required work relative to passive systems. In contrast, for strongly persistent motion, the control must increasingly counter-act the active motion, making transport to a given target position more costly. Nevertheless, positional feedback remains beneficial compared to open-loop control across the entire parameter range, including the limit of infinitely persistent activity.

Contrasting our results with earlier findings for direct measurements of the active degree of freedom itself~\cite{Schuttler_2025, GarciaMillan_2025},
we found that different measurement schemes exploit activity using two fundamentally different mechanisms. Direct measurements of the active fluctuations primarily harvest persistent motion itself, whereas position measurements predominantly exploit the increased potential energy due to active fluctuations, while only indirectly inferring their direction through steady-state correlations. This distinction provides a physical interpretation of why the two feedback strategies exhibit qualitatively different asymptotic behavior.

Since the present work relies exclusively on positional measurements, it broadens the applicability of the optimal control protocols to passive colloids immersed in active baths, where the microscopic active degrees of freedom are hidden from observation. More generally, our results demonstrate how information about accessible observables can be converted into energetic gains by exploiting nonequilibrium correlations.

We note that, while a realistic measurement generally suffers uncertainties, we throughout 
assumed error-free measurements. 
Extensions to noisy measurements are however readily possible following the approach in Refs.~\cite{GarciaMillan_2025, Abreu_2011ExtractingWork}. 

It would further be interesting to investigate more general active dynamics, such as run-and-tumble particles or active Brownian particles, which feature nonlinearities making the inference of hidden activity more involved~\cite{Frydel2022, Frydel2023, Tucci2022}. While Eqs.~\eqref{optimalSolution} remain valid for any exponentially decaying active model, possible nonlinear correlations between $x_0$ and $v_0$ may lead to qualitative different behaviors. Another promising direction is to combine multiple position measurements to infer more directly the stochastic velocity from positional measurements and quantify the resulting thermodynamic value of information in the spirit of \cite{muruga2026extractingworkhiddendegrees}. Finally, one could also include (simultaneous) control of the trap stiffness \cite{szamel2026machine,baldovin2023control,casert2024learning} to further enhance the output work. Beyond optimally transporting particles, these ideas may provide guiding principles for designing information engines and feedback protocols operating in active and driven nonequilibrium environments.

\section*{Acknowledgment}
We thank John Bechhoefer for valuable discussions on experimentally accessible measurements and inferred information. 
L.T.S. acknowledges support from the International Max-Planck Research School for Physics of Biological and Complex Systems fellowship program.

\appendix

\section{Derivation of the optimal work}\label{AppendixDerivation}

Here we derive the optimal work solutions given in Eq.~\eqref{optimalSolution} in the main text.
The main steps are analogous to the calculation in Ref.~\cite{Schuttler_2025}.

We start by identifying
\begin{align}
    W_{x_0}[\lambda] =  \int_0^{t_\mathrm{f}}\rmd s \dot\lambda(s)\kappa[\lambda(s) - q_{x_0}(s)]\,,\label{appW}
\end{align}
where $q_{x_0}$ is the average position conditioned on $x_0$ given by Eq.~\eqref{q_x0}. By rearranging Eq.~\eqref{q_x0} for $\lambda(s)$, the latter can be inserted in Eq.~\eqref{appW} after an integration by parts, yielding 
\begin{align}
    W_{x_0} =& \kappa\lambda(s)\left[\frac{1}{2}\lambda(s) - q_{x_0}(s)\right]_0^\tf \nonumber\\&+ \kappa\int_0^\tf\rmd s\, \dot q_{x_0}\left\{\frac{1}{\kappa}\left[\dot q_{x_0} - u_{x_0}\right] + q_{x_0}\right\}\,.
\end{align}
Note that we omit here and in the following the time arguments in the integrals which should be clear from context.
By simplifying the integral, we end up with
\begin{align}
   \!\! W_{x_0} =& \frac{\kappa}{2}[\lambda(s) - q_{x_0}(s)]^2\Big|_0^\tf + \int_0^\tf\rmd s \left[\dot q_{x_0}^2 - \dot q_{x_0}u_{x_0}\right].\label{appW2}
\end{align}
The temporal bulk contribution, i.e., the integral, is minimized by the solution to the Euler-Lagrange equation
\begin{align}
    \ddot q_{x_0} = \frac{1}{2}\dot u_{x_0}\,.\label{ELE}
\end{align}
Using Eq.~\eqref{u_x0}, Eq.~\eqref{ELE} is solved by
\begin{align}
    q_{x_0}(s) = x_0 + c s + \frac{\tau}{2}v_{x_0}\left(1-\mathrm{e}^{-s/\tau}\right)\,.\label{q_x0solve}
\end{align}
Now, all that remains is to determine the constant $c$ by inserting Eq.~\eqref{q_x0solve} into Eq.~\eqref{appW2}
\begin{widetext}
    \begin{align}
        W_{x_0} = \frac{\kappa}{2}\left\{\left[\lf - x_0 - c\tf - \frac{\tau}{2}v_{x_0}\left(1-\mathrm{e}^{-\tf/\tau}\right)\right]^2-x_0^2\right\} + \underbrace{\int_0^\tf \rmd s \left(c + \frac{v_{x_0}}{2}\mathrm{e}^{-s/\tau}\right)\left(c - \frac{v_{x_0}}{2}\mathrm{e}^{-s/\tau}\right)}_{=\tf c^2 - \frac{\tau}{8}v_{x_0}^2\left(1-\mathrm{e}^{-2\tf/\tau}\right)}\,.\label{appW3}
    \end{align}
\end{widetext}
To evaluate the optimal $c$, we solve $\nabla_c W_{x_0}=0$ for $c$ and get
\begin{align}
    c^*_{x_0} =& \frac{\kappa}{\kappa\tf+2}d_{x_0}\nonumber\\
    =&\frac{\kappa}{\kappa\tf + 2}\left[\lf - x_0 - \frac{\tau}{2}v_{x_0}\left(1-\mathrm{e}^{-\tf/\tau}\right)\right]\label{optCappendix}
\end{align}
The role of $d_{x_0}$, or rather $d_{v_0}=\mathbb E_{x_0}[d_{x_0,v_0}| v_0]$, has been discussed thoroughly in \cite{Schuttler_2025, GarciaMillan_2025}, where $d_{x_0,v_0} = \frac{\kappa}{\kappa\tf+2}\left[\lf - x_0 - \frac{v_0\tau}{2}(1-\rme^{-\tf/\tau})\right]$ is the effective distance recovered by measuring $\alpha=(x_0, v_0)$ initially. Both $d_{x_0}$ and $d_{v_0}$ quantify the initial distance of the measured position of the AOUP and the target position of the trap $\lf-x_0$ 
minus the distance traveled, on averaged, due to the persistence, before $v$ decorrelates. The difference between $d_{x_0}$ and $d_{v_0}$ is that, for the former, exact knowledge is known about the initial position while we only statistically infer the average displacement the active fluctuations cause, while the latter quantifies the opposite case where we know the exact displacement due to the active velocity with an inferred initial position. While a seemingly small difference, the exact knowledge on the excitation of position or active fluctuation manifests strongly in the resulting optimal work, e.g., see Fig.~\ref{fig:fig2}(b).

Inserting Eq.~\eqref{optCappendix} into the work Eq.~\eqref{appW3}, the AOUP trajectory Eq.~\eqref{q_x0solve}, and $\lambda_{x_0}$ recovers the optimal quantities listed in Eqs.~\eqref{optimalSolution}.

\section{Passive system with constant drift}\label{passiveDrift}
In this section, we show the details regarding the optimal feedback solutions of tilted passive systems and relate this in more detail to the infinitely persistent active cases. 

Let $v_0\in\mathbb{R}$ be a Gaussian random number and consider a system with potential $V_\mathrm{eff}(x) = \frac{\kappa}{2}(x-\lambda)^2-v_0x$, then
\begin{align}
    \rmd x(s) = -\kappa[x(s) - \lambda(s)]\rmd s + v_0\rmd s + \sigma\rmd B_s\,,
\end{align}
with noise averaged dynamics
\begin{align}
    \dot r(s) = -\kappa [r(s) -\lambda(s)] + v_0\,.
\end{align}
Using Ref.~\cite{monter2026energyefficientcontrolinteractingmicroscopic}, the optimal work can readily be found to be
\begin{align}
    W_{\mathrm{eff}} &= \Delta V_\mathrm{eff} + \tf c^2\label{Weff}\,.
\end{align}
Here, $\Delta V = V_\mathrm{eff}[r(\tf), \lf] - V_\mathrm{eff}[r(0), 0]$ is the potential energy difference of the final and initial state. The optimal slope is found by setting $\nabla_c W_{x_0,\mathrm{eff}}=0$, yielding
\begin{align}
    c^*_\mathrm{eff} = \frac{\kappa(\lf-x_0)+v_0}{\kappa\tf+2}\,.\label{ceff}
\end{align}
Subsequently, the optimal work is then found by inserting Eq.~\eqref{ceff} into Eq.~\eqref{Weff}
\begin{align}
    W^*_{\mathrm{eff}} =& \frac{\kappa}{\kappa\tf+2}(\lf-x_0)^2-\frac{\kappa}{2}x_0^2 - \frac{\tf}{2(\kappa\tf+2)}v_0^2 \nonumber\\&- \frac{\kappa\tf}{\kappa\tf+2}v_0(\lf-x_0)\,.\label{effWork}
\end{align}

\subsection{Comparison to active-fluctuation measurements}\label{passiveMappingv0}
We first consider the case where $v_0$ is directly measured. Because we start in a stationary distribution, the inferred position is $x^\mathrm{eff}_{v_0} = v_0/\kappa$, which coincides with $\lim_{\tau\to\infty} \mathrm{cov}\,(x_0,v_0)/\mathrm{var}(v_0) v_0$ in the active case [see Eqs.~\eqref{covariances}]. The optimal slope therefore reads
\begin{align}
    c = \frac{\kappa \lf - \kappa\frac{v_0}{\kappa}+v_0}{\kappa \tf + 2} = \frac{\kappa}{\kappa \tf + 2}\lf\,,\label{optCappendixPas}
\end{align}
and the optimal work is 
\begin{align}
    W_{\mathrm{eff}, v_0}^* =& \frac{\kappa}{2}\left(\lf - x_0-\frac{\kappa\tf}{\kappa\tf+2}\lf\right)^2-\frac{\kappa x_0^2}{2}\nonumber\\&-\frac{v_0\kappa\tf\lf}{\kappa\tf+2}+\frac{\kappa^2\tf\lambda_\mathrm{f}^2}{(2+\kappa\tf)^2}\nonumber\\
    =&\frac{\kappa}{\kappa\tf+2}\lambda_\mathrm{f}^2-\lf v_0\,.
\end{align}
Hence, comparing to the infinitely persistent limit of an initial $v_0$ measurement, we find
\begin{align}
    \lim_{\tau\to\infty}W_{v_0}^*=\frac{\kappa}{\kappa\tf+2}\lambda_\mathrm{f}^2 - \lf v_0 = W_{\mathrm{eff}, v_0}^*\,.\label{v0WorkLimit}
\end{align}
Assuming that the tilt is a random variable with values drawn from a Gaussian centered around zero, then Eq.~\eqref{v0WorkLimit} matches Eq.~\eqref{v0WorkTauInfty} both in individual tilt realizations as well as in expectation, further supported by
\begin{align}
    \lim_{\tau\to\infty}\Sigma_{x_0x_0|v_0} &=\frac{D}{\kappa} \equiv \Sigma_{x_0x_0}^\mathrm{pas}\,,\label{passiveCov}
\end{align} 
where $\Sigma_{x_0x_0}^\mathrm{pas}$ is the variance of a passive Ornstein-Uhlenbeck process.

\subsection{Comparison to position measurement}\label{passiveMappingx0}
So far, we have compared the tilted passive system with the feedback after active-fluctuation measurement. In contrast to $v_0$-measurements, if we measure the initial position, the value of $v_0$ is not fixed, but still a random variable. Hence, let now $x_0\in\mathbb{R}$ be fixed and $\tilde v_0 = \lim_{\tau\to\infty}v_{x_0} = \frac{\kappa\omega^2}{D\kappa+\omega^2}x_0$ be the inferred constant drift. (This relation is obtained by choosing the appropriate Gaussian distribution for $v_0$; and its physical origin is discussed in the main text. Here, we simply used here to show the correspondence.) The optimal slope Eq.~\eqref{ceff} therefore reads 
\begin{align}
    c = \frac{\kappa (\lf - x_0) +\tilde v_0}{\kappa\tf + 2}\,.\label{TiltOptimalc}
\end{align}
Inserting this, we can calculate the optimal work
\begin{widetext}
\begin{align}
    W^*_{\mathrm{eff}, x_0} =& \frac{\kappa}{2}\left(\lf - x_0 - \frac{\kappa\tf(\lf-x_0)+\tilde v_0\tf}{\kappa\tf+2}\right)^2-\frac{\kappa x_0^2}{2}-\tilde v_0\tf\frac{\kappa(\lf-x_0)+\tilde v_0}{\kappa\tf + 2} + \tf\frac{[\kappa(\lf-x_0)+\tilde v_0]^2}{(\kappa\tf+2)^2}\nonumber\\
    =& \frac{\kappa}{2}\left(\frac{2(\lf-x_0)-\tilde v_0\tf}{\kappa\tf+2}\right)^2-\frac{\kappa x_0^2}{2}-\tilde v_0\tf\frac{\kappa(\lf-x_0)+\tilde v_0}{\kappa\tf + 2} + \tf\frac{[\kappa(\lf-x_0)+\tilde v_0]^2}{(\kappa\tf+2)^2}\nonumber\\
    =& \frac{2\kappa(\lf-x_0)^2}{(\kappa\tf+2)^2}-\frac{2\kappa \tilde v_0\tf(\lf-x_0)}{(\kappa\tf+2)^2}+\frac{\tilde v_0^2 t_\mathrm{f}^2\kappa}{2(\kappa\tf+2)^2}-\frac{\kappa x_0^2}{2}-\frac{\tilde v_0\tf\kappa(\lf-x_0)}{\kappa\tf + 2} - \frac{\tilde v_0^2\tf}{\kappa\tf+2} \nonumber\\&+ \tf\frac{\kappa^2(\lf-x_0)^2 + 2\kappa \tilde v_0(\lf-x_0)+\tilde v_0^2}{(\kappa\tf+2)^2}\nonumber\\
    =& \frac{\kappa}{\kappa\tf + 2}(\lf-x_0)^2 + \frac{\tilde v_0^2\tf}{\kappa\tf+2}\underbrace{\left(\frac{\tf\kappa}{2(\kappa\tf+2)}+ \frac{1}{\kappa\tf+2}-1\right)}_{=-\frac{1}{2}}-\frac{\kappa}{2}x_0^2-\frac{\tilde v_0\tf\kappa(\lf-x_0)}{\kappa\tf+2}\,.\label{appendixW1}
\end{align}
\end{widetext}
One can immediately see that the work Eq.~\eqref{appendixW1} scales quadratically in the initial position $x_0$. Hence, 
it is convenient to write this as a polynomial in $x_0$ using that $v_{x_0}\propto x_0$, giving rise to
\begin{align}
    W^*_{\mathrm{eff}, x_0} =& \frac{\kappa}{\kappa\tf+2}\lambda_\mathrm{f}^2 - \lf\left(\frac{2\kappa}{\kappa\tf+2}x_0 + \frac{\tilde v_0\tf\kappa}{\kappa\tf+2}\right) \nonumber\\&+ \left(-\frac{\kappa^2\tf x_0^2}{2(\kappa\tf+2)}-\frac{\tilde v_0^2\tf}{2(\kappa\tf+2)}+\frac{\tilde v_0x_0\tf\kappa}{\kappa\tf+2}\right)\,.\label{intWeq}
\end{align}
We focus on the coefficient of the term proportional to $x_0^2$ by inserting Eq.~\eqref{vgivenx} into Eq.~\eqref{intWeq} and reading off the $\mathcal{O}(x_0^2)$ term
\begin{align}
    &\frac{-\kappa^2\tf}{2(\kappa\tf+2)} - \frac{\omega^4\kappa^2\tf}{2(\kappa\tf+2)(D\kappa+\omega^2)^2}+\frac{\tf\omega^2\kappa^2}{(\kappa\tf+2)(D\kappa+2)}\nonumber\\
    =&\frac{-\kappa^2\tf(D\kappa+\omega^2)^2 - \omega^4\kappa^2\tf+2\kappa^2\omega^2\tf(D\kappa+\omega^2)}{2(\kappa\tf+2)(D\kappa+\omega^2)^2}\nonumber\\
    =&\frac{-\kappa^2\tf(D^2\kappa^2+2D\kappa\omega^2+\omega^4) - \omega^4\kappa^2\tf+2\kappa^2\omega^2\tf(D\kappa+\omega^2)}{2(\kappa\tf+2)(D\kappa+\omega^2)^2}\nonumber\\
    =&\frac{-\kappa^4\tf D^2+\kappa^2\tf\omega^4 - \omega^4\kappa^2\tf}{2(\kappa\tf+2)(D\kappa+\omega^2)^2}\nonumber\\
    =&\frac{-\kappa^4\tf D^2}{2(\kappa\tf+2)(D\kappa+\omega^2)^2}\,.
\end{align}
Hence, the work optimal passive work with ``inferred drift'' reads
\begin{align}
    W^*_{\mathrm{eff}, x_0} =& \frac{\kappa}{\kappa\tf+2}\lambda_\mathrm{f}^2-\frac{2\lf\kappa}{\kappa\tf+2}\left[1+\frac{\tf\omega^2\kappa}{2(D\kappa+\omega^2)}\right]x_0 \nonumber\\&- \frac{\kappa^4\tf D^2}{2(\kappa\tf+2)(D\kappa+\omega^2)^2}x_0^2\,.\label{PassiveWorkShiftedX}
\end{align}
It should be emphasized that the optimal passive work with constant inferred drift [Eq.~\eqref{PassiveWorkShiftedX}] \emph{always} benefits from a position measurement with subsequent average over all realizations thereof. This is not a trivial statement, as the additional tilt in the potential adds complexity to the system and measurements of the tilt give no benefit [see Eq.~\eqref{v0WorkLimit}]. 

The optimal work in Eq.~\eqref{optimalSolution} can be written as $W_{x_0}^* = b_0 + b_1 x_0 + x_2 x_0^2$. We now compare the coefficients in Eq.~\eqref{PassiveWorkShiftedX} to $b_0$, $b_1$, and $b_2$ in the limit $\tau\to\infty$. Firstly, $b_0$ trivially matches the $\mathcal{O}(x_0^0)$ term in Eq.~\eqref{PassiveWorkShiftedX}. Thus, we only focus on the latter two
\begin{align}
    &\lim_{\tau\to\infty}b_1\nonumber\\=&\lim_{\tau\to\infty}\left\{-\frac{2\kappa\lf}{\kappa\tf+2}\left[1+\frac{\tau}{2}\frac{\omega^2\kappa\tau}{D(\kappa\tau+1)+\omega^2\tau}\left(1-\mathrm{e}^{-\tf/\tau}\right)\right]\right\}\nonumber\\
    =& -\frac{2\kappa\lf}{\kappa\tf+2}\left[1 + \frac{\omega^2\kappa\tf}{2(D\kappa+\omega^2)}\right]
\end{align}
and
\begin{widetext}
    \begin{align}
        \lim_{\tau\to\infty}b_2&=\lim_{\tau\to\infty}\left\{\frac{\kappa}{\kappa\tf+2}\left[1+\frac{\tau}{2}\frac{\omega^2\kappa\tau}{D(\kappa\tau+1)+\omega^2\tau}\left(1-\mathrm{e}^{-\tf/\tau}\right)\right]^2-\frac{\kappa}{2}-\frac{\tau}{8}\frac{\omega^4\kappa^2\tau^2}{[D(\kappa\tau+2)+ \omega^2\tau]^2}\left(1-\mathrm{e}^{-2\tf/\tau}\right)\right\}\nonumber\\
        &= \frac{\kappa}{\kappa\tf+2}\left[1+\frac{1}{2}\frac{\tf\omega^2\kappa}{D\kappa+\omega^2}\right]^2-\frac{\kappa}{2}-\frac{1}{4}\frac{\tf\omega^4\kappa^2}{[D\kappa+ \omega^2]^2}\nonumber\\
        &=\frac{\kappa\left\{4(D\kappa+\omega^2)^2 + 4\omega^2\tf\kappa(D\kappa+\omega^2) + \tf^2\omega^4\kappa^2-2(\kappa\tf+2)(D\kappa+\omega^2)^2-\tf\omega^4\kappa(\kappa\tf+2)\right\}}{4(\kappa\tf+2)(D\kappa+\omega^2)^2}\nonumber\\
        &=\frac{\kappa\left\{ 4\omega^2\tf\kappa(D\kappa+\omega^2) + \tf^2\omega^4\kappa^2-2\kappa\tf(D\kappa+\omega^2)^2-\tf\omega^4\kappa(\kappa\tf+2)\right\}}{4(\kappa\tf+2)(D\kappa+\omega^2)^2}\nonumber\\
        &=\frac{\kappa\left\{  \tf^2\omega^4\kappa^2-2\tf D^2\kappa^3+2\kappa\tf\omega^4-\tf\omega^4\kappa(\kappa\tf+2)\right\}}{4(\kappa\tf+2)(D\kappa+\omega^2)^2}\nonumber\\
        &=\frac{-\tf D^2\kappa^4}{2(\kappa\tf+2)(D\kappa+\omega^2)^2}\,,
    \end{align}
\end{widetext}
which both match the passive counterparts in Eq.~\eqref{PassiveWorkShiftedX}.

\section{Quasistatic limit}\label{quasistatic}
Next, we consider the quasistatic limit $\tf\to\infty$ of the optimal solutions discussed in the main text. 
The optimal work Eq.~\eqref{OptimalWork} is minimized in the limit $\tf\to\infty$, where it takes the form,
\begin{align}
    \lim_{\tf\to\infty}\mathbb{E}_{x_0}\left[W_{x_0}^*\right] = -&\left(\frac{1}{2} + \frac{\omega^4\kappa\tau^3}{8[D(\kappa\tau+1)+\omega^2\tau]^2}\right)\nonumber\\&\times\kappa\mathbb{E}_{x_0}\left[x_0^2\right]\,. \label{quasistaticWork}
\end{align}

The first term in Eq.~\eqref{quasistaticWork} corresponds to the amount of extractable energy in a passive system.  
In the presence of active fluctuations additional work can be extracted. In the quasistatic limit, \emph{all} of the initial correlations of the active fluctuations can be harvested, resulting in the second term of Eq.~\eqref{quasistaticWork}. 
The initial jump, $\lim_{\tf\to\infty}\lim_{s\to0^-}\lambda_{x_0}^*(s)=x_0-v_{x_0}/2\kappa$, differs from the instantaneous repositioning of the trap to the particle position $x_0$, which one would expect for a passive system \cite{Abreu_2011ExtractingWork}. Instead, the trap moves to an intermediate position, extracting a fraction of the initial potential energy. Subsequently, the trap follows the particle to extract work from the active fluctuations, until the latter decorrelate. For $s\gg\tau$, the protocol becomes linear in time to minimize the work lost due to frictional resistance, consistent with the optimization of passive systems \cite{Schmiedl2007, monter2026energyefficientcontrolinteractingmicroscopic}.

\section{Scaling of the variance of the active fluctuations}\label{AppendixScaling}
The scaling of $\omega^2$ in terms of $\tau$ impacts the results of the optimization shown in this manuscript, in particular, the asymptotic limits. As mentioned in the main text, the choice we use, $\omega^2\sim\tau^0$, enables us to compare directly with previous work~\cite{Schuttler_2025, GarciaMillan_2025}, wherein the same scaling was used to equate the fluctuations of an AOUP and a run-and-tumble particle (RTP) up to the second moment, as well as yielding a steady state for all values of $\tau$. 

However, it is \emph{a priori} unclear if this is the most reasonable choice to make. For instance, it implies that the noise amplitude depends on the persistence time-scale $\sigma'\sim\tau^{1/2}$. Another scaling choice could be, for example,  $\sigma'\sim\tau^0$, which would imply $\omega^2\sim \tau^{-1}$. Changing the scaling has crucial implications. For instance, the ``passive limit'' $\tau\to0$ does not reduce to a passive system with temperature $D$, but rather to a system with temperature $D+D'$ with $D'={\sigma'}^2/2$. Such a comparison between small persistence active systems and passive systems brings up questions like ``what temperature do we compare with?''. Moreover, 
the optimal average work with this alternative scaling becomes
\begin{align}
 	\tilde{\mathcal W}^\mathrm{ex}_\mathrm{act-pas,x_0} =& \frac{\kappa \tf}{2(\kappa\tf + 2)}(D_\mathrm{pas} - D) + \frac{D'\tf}{2(\kappa \tf + 2)} \frac{1}{\tau} \nonumber\\&+ \mathcal{O}\left(\tau^{-2}\right)\,.
\end{align}
Thus, with an appropriate choice for the temperature $D_\mathrm{pas}$ of the passive comparison, i.e., $D_\mathrm{pas}=D$, the excess feedback work then approaches zero from above in the infinitely persistent limit $\tau\to\infty$. Note that this choice $D_\mathrm{pas}=D$ immediately implies $\lim_{\tau\to0}\tilde{\mathcal W}^\mathrm{ex}_\mathrm{act-pas,x_0}<0$. Hence, there exists again a region of the parameter space, where activity is beneficial and a complementary region where activity leads to a positive excess work, respectively.

\bibliography{sources}
\end{document}